\documentclass[aps,prb,twocolumn,superscriptaddress,nofootinbib,showkeys,floatfix]{revtex4-2}

\usepackage{bm}
\usepackage{graphicx}
\usepackage{amsmath}
\usepackage{booktabs}
\usepackage[colorlinks=true,linkcolor=blue,citecolor=blue]{hyperref}

\begin{document}
\setcounter{dbltopnumber}{1}
\title{Auditing a superconducting database built by a large language model}
\author{Wenjian Lu}
\thanks{Corresponding author:
\href{mailto:wjlu@issp.ac.cn}{wjlu@issp.ac.cn}}
\affiliation{%
Key Laboratory of Materials Physics, Institute of Solid
State Physics, HFIPS, Chinese Academy of Sciences, Hefei 230031, China}
\date{\today}

% The columns are not stretched to equal height.  The first-page footnote
% shortens the left column, and with the columns balanced the text of
% the right one is pushed away from the text of the left.
\raggedbottom

\begin{abstract}
The data on superconducting materials are published as measurements and
scattered across the literature, in prose, in tables and in figures, so
that gathering them into one place means reading the papers themselves.
Large language models (LLMs) can read those papers, and databases built
that way are appearing.  The difficulty is not the reading but the
trust: a number lifted out of a sentence can belong to a neighbouring
compound, a computed transition temperature can be recorded without
saying that it is computed, and the paper itself may be wrong.  This
paper describes a pipeline that extracts superconducting materials from
the published literature with an LLM, checks every value against its
source, and asks how far that checking goes.  The pipeline proposed
19{,}576 records; a cascade of gates
rejected 659{,}252 field-level extractions, 644{,}438 of them for want
of verbatim evidence, 13{,}121 because the text does not attach the
value to the material, and 1{,}693 because the source sentence reads
as a calculation.  The database that survives holds 3{,}137 material
records, each with the sentence its value came from.  An audit of the
survivors by two independent models, on 500 records drawn in
proportion to the database, shows that the gates remove the errors that
are mechanical: in 93.4\% of the audited records the value is in the
text and the text is about the compound the record names, and 91.8\%
of them are measurements rather than predictions.
\end{abstract}


\maketitle

\section{Introduction}

Superconductivity is the phenomenon in which the resistance of a
material vanishes below a transition temperature $T_c$, and the
measurements are published in the
papers -- from the first observation \cite{onnes} onwards.  The consequence is that the data on superconductors are
scattered across journals, in prose, in tables and in figures, and the
compilations that gather them are small.  The NIMS SuperCon database
collects transition temperatures by hand \cite{supercon}; the
Crystallography Open Database holds the structures \cite{crystallogra2011};
the Materials Project supplies computed properties for millions of
compounds \cite{commentaryth2013}.  Each is valuable and each is
bounded by the work that went into it.

The literature itself is not bounded that way.  The extraction of
materials and their properties from published text has a long history
\cite{materialssyn2017}, and the arrival of LLMs has turned it into an
automated task: models read papers and return structured records, and
databases are now being assembled that way for magnetic
\cite{gptarticleex2024,thenortheast2025} and thermoelectric
\cite{largelanguag2025} materials.  The superconductivity literature has
been mined in the same way \cite{automaticext2023,superconduct2022}, and
a curated dataset of superconductors with crystal structures has
recently been assembled \cite{3dscadataset2023}.

The weakness of the approach is the same in every domain, and it is not
a weakness of the models: it is a weakness of trusting them.  A model
that reads a paragraph will report the transition temperature of the
compound next door, or the value from a first-principles paper without
saying that it is computed, or a number that the paper never states.  The obvious remedy is
to check each value against its source, and the obvious next question
is how well that works.  The answer cannot be read off the pipeline:
the pipeline only ever sees what it kept.  Where an LLM-built database
has been checked at all, the check has been a second LLM reading the
same sentences \cite{thenortheast2025} -- which is a useful thing to
do and not the same as a check by someone who knows the field.

This paper describes a pipeline that extracts superconducting materials
from the published literature with an LLM, and then measures how
trustworthy what it kept is.  Three things make the measurement
possible.  The first is a cascade of checks -- verbatim evidence,
material binding, physical plausibility, how the value was obtained, and
agreement across papers -- applied to every extracted
value rather than to every paper.  The second is that the database
keeps the sentence each value came from, so any record can be checked
against its source by a reader, by another model, or by an expert,
without re-running the extraction.  The third is an audit with two
judges, neither of them the model that did the extraction, and each of
them asked the same three questions about each value.

\section{The pipeline}

\begin{figure}[!t]
\includegraphics[width=\columnwidth]{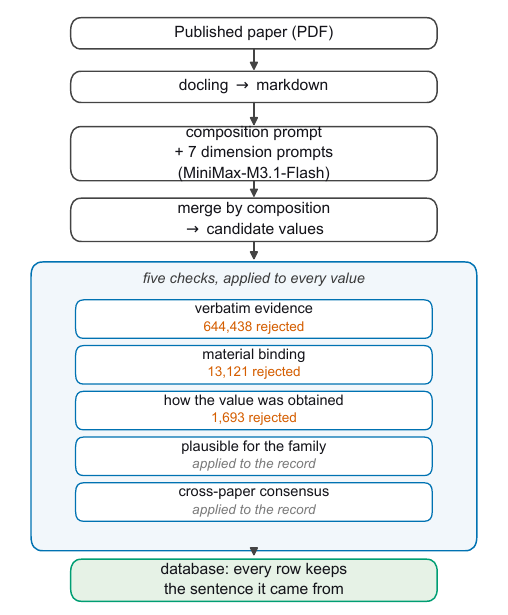}
\caption{The pipeline and the five gates.  The counts are the
field-level extractions each gate rejected, over the extraction of the
corpus; the two gates that reject nothing at the field level are
applied to the record as a whole.}\label{fig:pipeline}
\end{figure}

\subsection{From PDF to records}

Papers are collected from the publishers, converted to text with
docling \cite{doclingconve2026}, and read by an LLM
(MiniMax-M3.1-Flash) through one prompt for the compositions and seven
dimension-specific prompts rather than through one: base
superconducting properties, pressure dependence, isotope and Uemura
analysis, specific heat, Fermi parameters and resistivity, crystal
structure, and preparation.  The dimension split matters for what
follows, because it lets the evidence check be applied per field: a
paper that reports $T_c$ but no $H_{c2}$ is a paper from which $T_c$
can be taken and $H_{c2}$ cannot.

The dimension results are merged into one record per composition, and a
record carries the composition, the superconducting properties, and the
crystalline and preparation data where the paper gives them.  Nothing
is written to the database yet.

\subsection{The five gates}

\begin{table}[t]
\caption{Field-level extractions rejected by the gates, over the
extraction of the corpus.  The counts are of individual values, not of
records: one record offers up to ten properties.  Every paper is
counted once, from the run whose result reached the database; a paper
that was extracted more than once is counted from its last run.
\label{tab:gates}}
\begin{ruledtabular}
\begin{tabular}{lrl}
\toprule
Field & Rejected & Reason\\
\midrule
$T_{\rm CDW}$                 & 79{,}323 & no verbatim evidence\\
$H_{c1}$                      & 79{,}341 & no verbatim evidence\\
$J_c$                         & 78{,}490 & no verbatim evidence\\
Penetration depth             & 78{,}233 & no verbatim evidence\\
Coherence length              & 77{,}207 & no verbatim evidence\\
$H_{c2}$                      & 76{,}419 & no verbatim evidence\\
Energy gap                    & 75{,}955 & no verbatim evidence\\
$T_c$                         & 53{,}579 & no verbatim evidence\\
Pairing symmetry              & 23{,}213 & no verbatim evidence\\
Mechanism                     & 22{,}678 & no verbatim evidence\\
$T_c$                         & 13{,}121 & not stated for this material\\
$T_c$                         &  1{,}693 & a calculation\\
\midrule
Total                         & 659{,}252 & \\
\bottomrule
\end{tabular}
\end{ruledtabular}
\end{table}

% Declared here, and not in the section that discusses it: a float that
% spans both columns can only go to the top of the page after the one it
% is declared on, so this is what puts the corpus figure at the top of
% the page the database section starts on.
\begin{figure*}[!t]
\includegraphics[width=\textwidth]{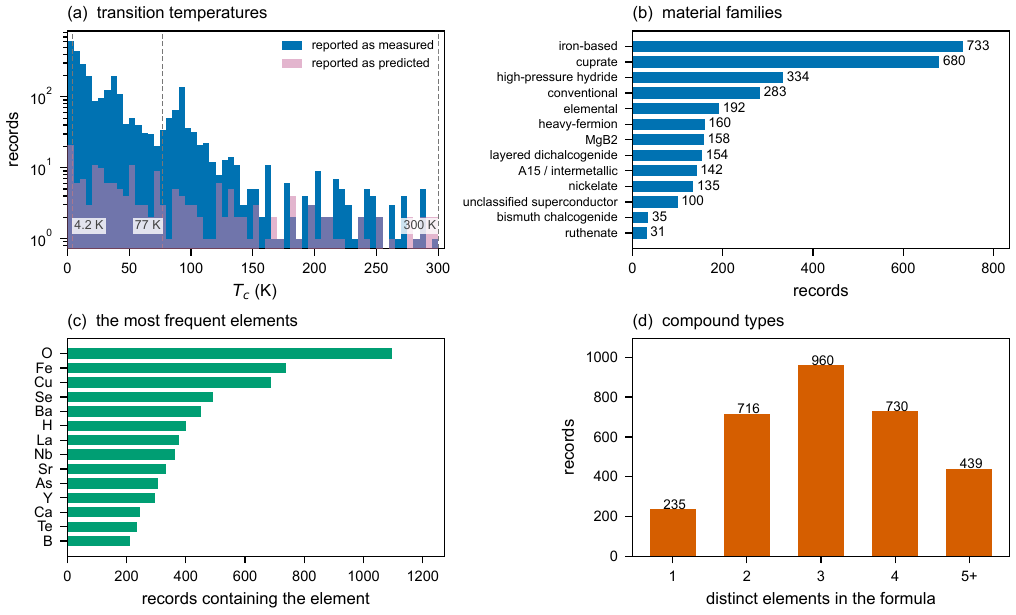}
\caption{The corpus of extracted superconducting materials.  (a) The
distribution of transition temperatures, with the records whose source
reports the transition temperature as a prediction drawn translucently
over those it reports as an observation; the dashed lines mark the temperatures of
liquid helium (4.2~K), liquid nitrogen (77~K) and room temperature
(300~K).  (b) The material families.  (c) The most frequent elements.
(d) The number of distinct elements in the formula.  The count is
3{,}137 records over 1{,}371 distinct compositions, from a corpus of
26{,}210 papers (21 November 2026).}\label{fig:database}
\end{figure*}

A value that survives the merge is a candidate, and every candidate is
put through five checks.

\emph{Verbatim evidence.}  The number must appear in the paper,
spelled the way the paper spells it.  This check fails more often than
any other, and it fails for a reason that has nothing to do with the
model: most properties are not in most papers.  The comparison folds
typographic variants, so that ``$T_c = 39\,$K'' and ``39 K'' are the
same number, but ``39 K'' is still not ``39.5 K''.

\emph{Material binding.}  The sentence that carries the value must also
name the material, or a normalization of it.  Superconductor papers
name many compounds in one paragraph, and a model reading it will
attach a neighbouring compound's $T_c$ to the subject of the sentence.
This check is what stops ``FeSe'' from being recorded as 109~K, the
transition of a monolayer on a substrate, reported in the same
paragraph as the bulk transition.

\emph{Physical plausibility.}  Every family has a documented range of
accepted transition temperatures, and a value outside it is rejected.
The check is deliberately conservative: rejecting a real measurement
costs one record, while accepting a misattributed one corrupts the
database silently.

\emph{How the value was obtained.}  Papers often quote both a
predicted and an observed transition temperature in the same paragraph,
and the headline theoretical value wins.  The classifier re-reads the
sentence around the value: wording of prediction sends the value out,
wording of observation keeps it.  Where the wording settles the
question the record follows it; where the sentence carries no wording
either way, the record keeps the extraction model's own judgement of
what the number is, and that judgement is what the flag in the database
records.

\emph{Cross-paper agreement.}  A compound reported by several papers
carries several values, and no single check can catch a measurement
that is the paper's own but wrong for that material.  Values within a
tolerance of one another are collapsed onto their cluster; where no
cluster dominates, the better-documented record is kept.

\subsection{What the gates reject}

Table~\ref{tab:gates} counts what each gate threw away.  The pattern is
the point: the rare properties are rejected far more often than the
common one.  $T_{\rm CDW}$, $H_{c1}$, $J_c$, the penetration depth, the
coherence length, $H_{c2}$ and the energy gap each lost more than
75{,}000 field-level extractions, while $T_c$, the property that nearly
every superconductor paper states, lost 53{,}579.  Pairing symmetry
and mechanism, which are reported in words rather than in numbers, lost
fewer still.

The three reasons in the last two rows are worth separating, because
they are different failures.  A value with no verbatim evidence was
probably never in the paper.  A value that the text does not attach to
the material was read from the wrong place: the sentence named a
different compound.  A value that the text calls a calculation is not
a failure of reading at all: the paper states the number correctly and
says what it is.  Such a value is dropped when the sentence carrying
it reads as a calculation, which is what these 1,693 are; the computed
values that do survive are the ones whose sentence carries no wording
either way, and there the record keeps the extraction model's own
judgement of what the number is, so that the distinction between the two
kinds of number survives into the database.
Figure~\ref{fig:pipeline} shows the cascade and the counts that reach
the table.

\section{The database}

The corpus is a snapshot of 26{,}210 papers.  The extraction has been
run over 26{,}184 of them and proposed 19{,}576 records; 8{,}076 of
those named a composition that is not a superconductor, which is the
characteristic hallucination of a model reading a reference list.  Of
the 11{,}500 that survived that filter, 7{,}090 became new rows, and
the rest were merged into rows that already existed or were dropped
because their own evidence sentence showed a calculated transition
temperature.  Consolidation across papers and deduplication leave
3{,}137 records over 1{,}371 distinct compositions, each with the
sentence its value came from.

Figure~\ref{fig:database} shows what the corpus holds.  The transition
temperatures run from 0.01~K to 298~K, four and a half decades, of which
five records sit at 0~K, and the distribution is the one the literature
gives: a dense
band below 30~K and a long tail of high-$T_c$ compounds.  The families
are the familiar ones (iron-based and cuprate materials dominate, with
a substantial hydride family), and 100 of the records are still
unclassified, which is a fact about the extraction rather than about
the physics.  Just over half of the compounds are
binary or ternary.

\begin{figure}[!t]
\includegraphics[width=\columnwidth]{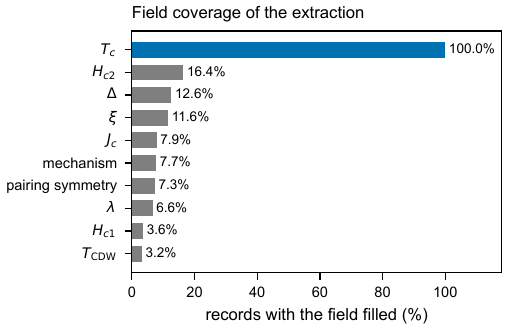}
\caption{How much of each field the extraction actually fills.  The
fraction of the 3{,}137 records in which the field is present.}\label{fig:coverage}
\end{figure}

The fields are not equally well covered, and the coverage measures the
extraction rather than the literature.  Figure~\ref{fig:coverage} shows
the fraction of records in which each property is filled.  $T_c$ is
present almost everywhere, because it is the one property that nearly
every paper in the corpus states.  $H_{c2}$, the coherence length, the
penetration depth and the energy gap are present for a minority, and
pairing symmetry and mechanism, which are reported as words and to

which the verbatim check cannot be applied directly, are the
rarest.

\section{Auditing the survivors}

\begin{table}[!t]
\caption{The two audits of the surviving records.  Both judges were
asked the same three questions, so their rates are comparable, and the
text compares them record by record.  Rates carry 95\% Wilson
intervals.  The third question asks how the value was obtained, not
whether the record stands: a computed value that the paper states for
the right material is a correct record.}\label{tab:audit}
\begin{ruledtabular}
\begin{tabular}{llcc}
\toprule
 & \textbf{question} & \textbf{rate} & \textbf{95\% interval}\\
\midrule
First & value is in the text & 96.8\% & 94.9--98.0\%\\
First & text is about the material & 94.6\% & 92.3--96.3\%\\
First & both & 93.4\% & 90.9--95.3\%\\
First & value is a measurement & 91.8\% & 89.1--93.9\%\\
Second & value is in the text & 95.2\% & 93.0--96.8\%\\
Second & text is about the material & 93.8\% & 91.3--95.6\%\\
Second & both & 91.8\% & 89.1--93.9\%\\
Second & value is a measurement & 90.4\% & 87.5--92.7\%\\
\bottomrule
\end{tabular}
\end{ruledtabular}
\end{table}

The gates are only as good as the question they ask, and the question
they ask is whether a value is stated in the paper for the material.
It is not whether the paper is right.  A value that the paper states
as computed is a record that stands: it is not an error, and it is not
an observation either, and the database keeps it with that fact
recorded.  Three
examples from the
pipeline's own history show the difference.  The monolayer
FeSe/SrTiO$_3$ transition, 109~K, was recorded for bulk FeSe, whose
transition is near 8~K: the sentence did state 109~K and it did name
FeSe, but not in the sense the record meant.  MgB$_2$'s 39~K was
recorded for YBa$_2$Cu$_3$O$_{7-\delta}$, from a paragraph that gave
both.  And the hydrides entered the database at the values of
first-principles studies (LaH$_{10}$ at 288~K and YH$_9$ at
276~K), with a single pressure point of 56~K written to four
compounds at once.

To measure how much of this survives, a stratified sample of the
corpus was audited twice: by two models that had no part in building
the database, each of them reading the whole paper the value came
from.  The
sample is 500 records, drawn by material family and by
transition-temperature band with the allocation in proportion to the
corpus, so that the rates it gives are the rates of the database and
not of a sample that over-weighs the rare families; every family and
band that is not empty is in it, and the draw is seeded.  The records
whose compounds the pipeline already knew to be troublesome are
marked, and they are reported apart: the hard records agree with the
first judge less often than the rest (90.6\% against 94.5\%).

% Declared here, and not at the head of the section: a float that is
% put at the top of a column takes the first column that is still
% open, and the table above has already taken the left one.  From
% here it goes to the top of the right-hand column of this page.

\subsection{The two model audits}

% Declared here, at the head of the subsection that discusses it: a
% float that spans both columns goes to the top of the page after the
% one it is declared on, and this is what puts the audit figure at the
% top of the page this subsection ends on.
\begin{figure*}
\includegraphics[width=\textwidth]{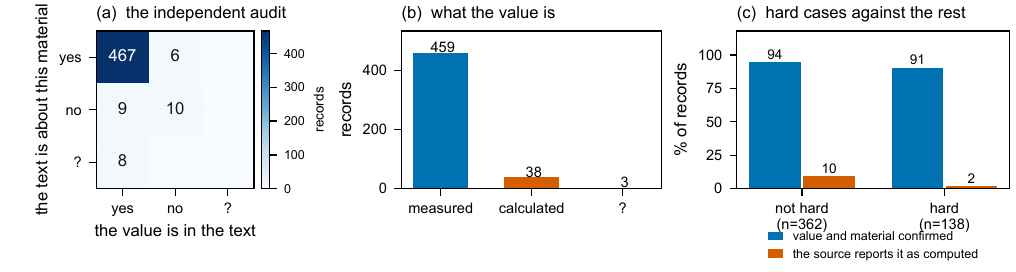}
\caption{What the first audit found.  (a) The value in the text against
the text being about the material; the first judge confirmed both in
467 of the 500 records it judged.  (b) What the extracted values are:
most are measurements, and 7.6\% are values the source paper reports as
computed.  (c) The same two rates
separated into the records the sample was built to be hard and the
rest.}\label{fig:audit}
\end{figure*}

The sample was audited twice, by two models that had no part in
building the database.  Each was given the whole paper the value came
from, not a window around the value, and each was asked the same three
questions: is this value in the text, is this text about this material,
and is the value a measurement or a calculation.  The third question
describes how the value was obtained and does not decide whether the
record stands: a computed value that the paper states for the right
material is a correct record.  The first is
DeepSeek-V4.1-Flash and the second DeepSeek-V4-Pro.  The whole paper
is what makes the answers mean something: a
judge that sees only the sentence that the value came from cannot tell
whether the material is named further on, and it rejects records that
are right.

Of the 500 records the first judge read, the value was stated in the
text in 484 cases (96.8\%, 95\% Wilson interval 94.9--98.0\%), and the
sentence was about the material the record names in 473 (94.6\%,
92.3--96.3\%).  The two conditions hold together in 467 records
(93.4\%, 90.9--95.3\%), and that is the rate the rest of this paper
uses.  The hard cases are where the first judge is least reliable: the
two conditions hold together in 125 of the 138 (90.6\%, 84.5--94.4\%)
against 342 of the 362 others (94.5\%, 91.6--96.4\%); the second judge
separates them more sharply, 120 of the 138 (87.0\%, 80.3--91.6\%)
against 339 of the 362 others (93.6\%, 90.6--95.7\%).

Thirty-three records the first judge does not accept.  Read by hand
they fall into two kinds.  Sixteen are values the judge cannot find in
the text: a family's transition stored on one of its members (BaFe$_2$As$_2$
at 29~K, where the paper gives the figure for the '122' family; ZrNCl
at 26~K, for the layered nitrides; CeIrIn$_5$ at 2~K, for the Ce-115
family), a value read from a sample that is not the compound (Nb at
0.55~K, from a junction; Nb$_3$Sn at 9~K, from a cavity), an increase
stored as a temperature (MgB$_2$-GaP at 0.8~K), a bound stored as a
value (CuH$_2$ at 40~K, where the paper writes ``$T_c > 40$~K''), a
value the paper computes rather than measures (Cs$_x$MoS$_2$ at 5~K,
whose sentence fixes $T_c$ by tuning a coupling constant), and a
statement of the opposite kind (FeSe at 6.5~K, from a sentence saying
the films were not superconducting down to 6.5~K).  The other seventeen
are records whose value is in the text but whose sentence the judge
does not take to be about the material the record names; seven of them
are ones where it cannot decide whether the sentence names the
material at all.

The third question asked how the values were obtained.  7.6\%
(5.6--10.3\%) of the audited records are, in the first judge's reading,
values that the source paper reports as computed (Fig.~\ref{fig:audit}),
and the second judge puts the share at 7.4\% (5.4--10.0\%).  That is
close to the 5.9\% the corpus itself carries, and the two judges pick
out the same records: the 23 computed values the corpus also marks as
computed are the same 23 for both of them, and neither judge finds a
measured value that the corpus has marked as computed.  Reading the
computed records by hand confirms them: ScH$_6$ is recorded at 100~K
from a sentence that reads ``We calculate $T_c$ with typical values of
$\mu^*$ such as 0.10 and 0.15, and obtain high critical temperatures of
around 100 K'', and LiCaH$_{12}$ at 288~K from a list of computed
hydrides and the values they are predicted to reach.  A value the
paper computes and states, for the compound the record names, is a
faithful record of what the paper says, and it is the record's mark
that says it is computed rather than measured.  The
database keeps these values and marks them, so that a reader can ask
for measured temperatures alone, or for the predictions as well.

The second judge reaches the same answer as the first on 470 of the 500
records: 453 that both accept, 27 that neither accepts, and 20 that
they judge differently (Table~\ref{tab:audit}).  Where the two judges
differ, the disagreement is about the evidence
and not about the value.  On FeSe at 70~K the first judge accepts the
record and the second rejects it; on BaFe$_2$As$_2$ at 29~K and on
FeSe$_0.83$S$_0.17$ at 10~K the second accepts what the first
rejects; on (Li$_1-x$Fe$_x$)OHFeSe at 41.9~K and on YBa$_2$Cu$_3$O$_{7-\delta}$
at 92~K it is the first that accepts what the other does not.  On the
first two questions alone the two judges differ on thirty records, and
on the third on thirteen.  The agreement between them is therefore a fact
about the two models and not a guarantee about the records.

% Declared at the end of the page before the section that discusses
% it: a float that spans both columns goes to the top of the page
% after the one it is declared on, and this is what puts the figure
% at the top of the page that section starts on.
\begin{figure*}
\includegraphics[width=\textwidth]{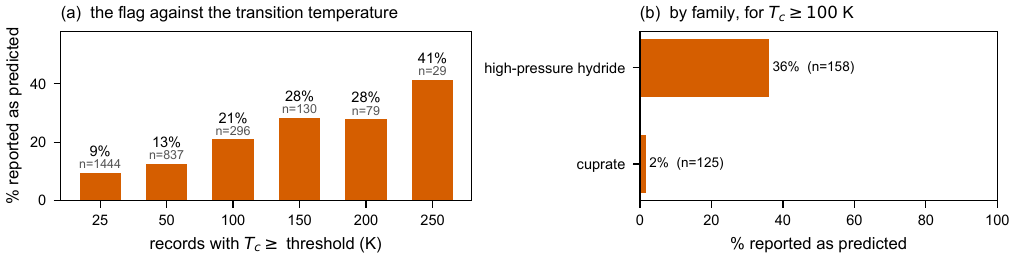}
\caption{Where the records whose source reports a prediction sit.  (a)
The fraction of records whose source reports the transition temperature
as a prediction, among those whose
transition temperature reaches the threshold.  (b) The same fraction by
family, for the records above 100~K; families with fewer than ten such
records are omitted.  The flag is the extraction's own reading of the
source paper, and it is set per record, not per value.}
\label{fig:tail}
\end{figure*}

\section{What the top of the distribution is made of}

The audit puts the share of computed values in the corpus at 7.6\%, and
the second judge at 7.4\% --- close to the 5.9\% the corpus itself
carries.  A share like that is easier to use if it can be
located, and the database carries the information needed to locate it:
every record keeps a flag for whether its source reports the transition
temperature as an experimental observation or as a prediction, and every
record keeps the sentence that the value was taken from.

Figure~\ref{fig:tail} uses that flag over the whole corpus.  Of the
3{,}137 records, 185 (5.9\%) are flagged as not experimental, but the
flag is not spread evenly.  It marks 9\% of the records from 25~K
upwards, 13\% of those from 50~K, 21\% of those from 100~K, 28\% of
those from 150~K and 41\% of those from 250~K.  The families differ by
an order of magnitude: of the 158 hydride records above 100~K, 57
(36\%) are flagged as predicted, and of the 125 cuprate records above
100~K, 2 (2\%) are.

The flag is the extraction's own reading of its source, and the audit
is a second reading of the same records by a model that had no part in
the extraction.  Set against each other on the 500 audited records,
the two readings disagree in one direction
only: of the 38 records the first judge calls computed, the flag calls
23 of them computed as well, and none of the 459 records the judge calls
measured is marked as computed by the flag.  Fifteen computed values are
therefore stored as measured.  Every share the flag
produces is a lower bound on the share of computed values in the corpus,
and the audit's 7.6\% is the better estimate of
what the corpus contains.

The highest transition temperatures in the database are therefore a
mixture of two kinds of statement.  The two records at the top are
both hydrides at 298~K, and they are of two different kinds:
LaSc$_2$H$_{24}$, whose sentence records that superconductivity was
confirmed both by zero resistance and by the suppression of $T_c$
under an external magnetic field, and Sr$_2$Sc$_4$H$_{33}$, whose
sentence is a structure label from a paper on predicted structures.
Four of the five records above 290~K are computations.

That is a physical question the corpus can be asked and answered, and
it is the reason the database is worth keeping in this form.  A search
for a high-$T_c$ superconductor returns these records first, and a
reader who is looking for a material to make and a reader who is
looking for a prediction to aim at are served by different ones.  Here
the difference is recorded per record, and any reader can check it
against the sentence that the value came from.

% Declared at the end of the section that precedes it, so that the
% figure lands at the top of the page this section starts on.

\section{What the database shows}

% Declared at the head of the section that discusses it: a float
% that spans both columns goes to the top of the page after the one
% it is declared on, and this is what puts the figure at the top of
% the page the section starts on.
\begin{figure*}
\includegraphics[width=\textwidth]{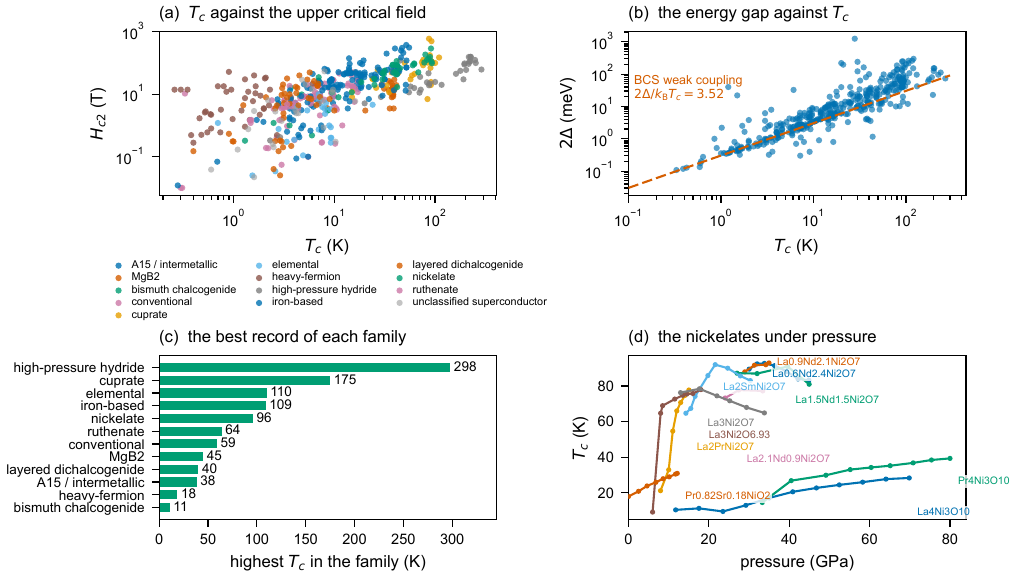}
\caption{Four relations.  (a)--(c) are drawn from the surviving records:
(a) The
transition temperature against the upper critical field, on logarithmic
axes, coloured by family.  (b) The energy gap against the transition
temperature, with the weak-coupling limit of the BCS relation
\cite{bc}, $2\Delta/k_{\rm B}T_c = 3.52$, for reference.  (c) The highest
transition temperature recorded in each family.  (d) The dependence of the
superconducting transition temperature of the nickelates on
pressure.}\label{fig:relations}
\end{figure*}

The database is not only a record of the literature; it is a corpus on
which the relations of the field can be looked at.  Figure~\ref{fig:relations}
draws four of them from the surviving records.  They are the figures the
project website draws, redrawn from the same data so that they can be
printed.

The first two are the checks that any compilation of superconductors
has to pass, and the database passes them the way the literature does.
The upper critical field rises with the transition temperature, across
the three decades of $T_c$ the corpus covers --- the correlation the
McMillan formula is built on --- and the hydrides sit at the high-$T_c$
end of it, with fields up to 158~T, although the largest fields in the
corpus are not theirs: a cuprate reaches 600~T at 87~K and the
iron-based compounds 377~T (Fig.~\ref{fig:relations}a).  The energy gap
follows the transition temperature over three decades as well, and the
records lie on both sides of the weak-coupling line: 76\% of
them above it, and the ones furthest below mostly the materials with the
smallest transition temperatures (Fig.~\ref{fig:relations}b).  The tail
of that scatter is not gaps at all: the extracted column mixes
quantities, and a handful of records carry a value that belongs to
something else.

The third panel is the ranking a database naturally produces, and it is
the ranking the audit is about.  Two of the records that head their
family are the errors this paper has already shown: the cuprate's
175~K is the number ``$T_c = 0.15$ corresponds to about 175~K'', a
scaling result stored as a transition temperature, and the
iron-based 109~K belongs to a FeSe monolayer and was recorded for
bulk FeSe.  A third, the MgB$_2$ record at 45~K --- MgB$_4$P$_4$, the
highest record in that family --- is marked by the extraction itself as
computed.  A reader who took the ranking at face
value would put a cuprate above every measured cuprate.  The record
that heads the hydride family, 298~K, is the one whose sentence
reports a measurement.

The fourth panel is what the field would otherwise have had to read
paper by paper: the pressure dependence of the nickelates, eleven
curves, the La$_3-x$Nd$_x$Ni$_2$O$_7$ domes between 6 and 45~GPa, the
two triple-layer compounds running out to 70 and 80~GPa, and the
highest transition temperature among them 93~K
(Fig.~\ref{fig:relations}d).  These curves are what the SuperCon
companion set contributes, and they are the clearest case for a corpus
that holds every point of a measurement rather than one number per
compound.

\section{Methods}

\subsection{The corpus and the documents}

Papers are collected from the publishers through an unattended download
loop that queries the literature by journal, by author and by material
name, and that keeps a content-hashed cache of the PDFs it has
already seen.  The documents are converted to text with docling
\cite{doclingconve2026}, and the markdown that the conversion
produces is what every later step reads.  A paper that cannot be
converted, and a paper whose text holds no superconducting quantity at
all, contribute nothing.

% Declared at the end of the page before the section that discusses
% it: a float that spans both columns goes to the top of the page
% after the one it is declared on, and this is what puts the
% screenshot at the top of the page that section starts on.

\subsection{The extraction}

The text is handed to an LLM (MiniMax-M3.1-Flash) through one prompt
for the compositions, with their experimental flag, and seven
dimension-specific prompts: the base superconducting properties, the
pressure dependence, the isotope and Uemura analysis, the specific
heat, the Fermi parameters and the resistivity, the crystal structure,
and the preparation.  Three of the seven are only asked when the paper
is about their subject --- an isotope effect is asked for when the text
mentions isotopes, a Fermi surface when it mentions Fermi surfaces ---
while the other four are always asked, so a paper of the kind that
describes no isotope effect is not billed for one.  The answers are
merged into one record per composition, keyed by the composition
itself, so that a property found by one prompt and a property found by
another end up on the same row.

The extraction of the corpus cost 115{,}435 calls to the model and
$7.3\times10^{8}$ tokens, of which $1.0\times10^{8}$ were completion
tokens and 30\% of those reasoning tokens.  The cost is that of the
reading; the cost of checking what the reading produced is what the
rest of this paper measures.

% Declared at the end of the page before the section that discusses
% it: a float that spans both columns goes to the top of the page
% after the one it is declared on, and this is what puts the
% screenshot at the top of the page that section starts on.
\begin{figure*}[!t]
\includegraphics[width=\textwidth]{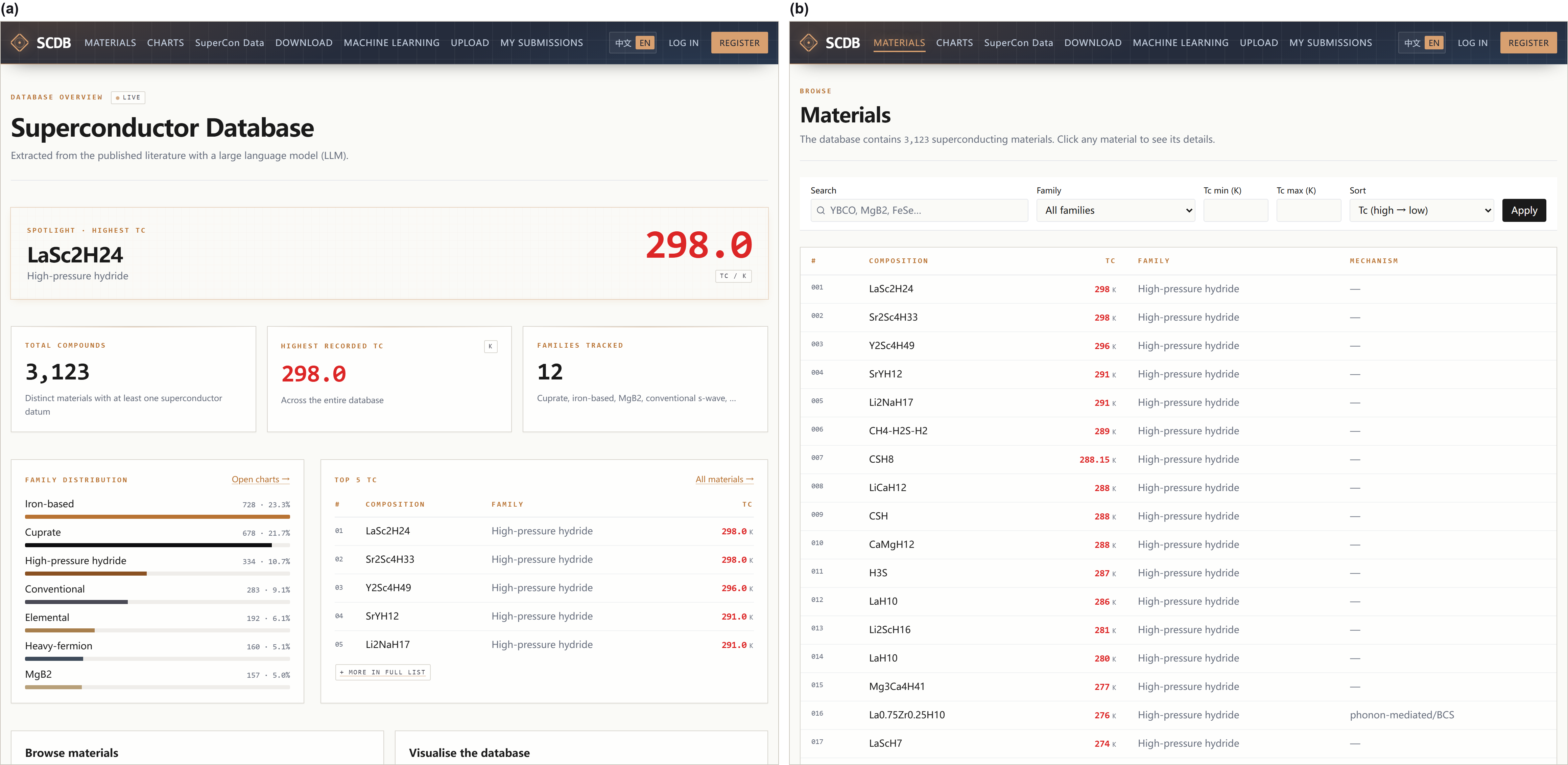}
\caption{The database as a website.  (a) The home page, which gives the
size of the corpus, its highest record and the families it holds.  (b)
The materials page, where the records are listed and can be narrowed by
composition, family, transition-temperature range and sort order.  The
captures were made in English; the site also serves Chinese.}\label{fig:site}
\end{figure*}

\subsection{The gates}

Four checks are applied to the merged records, and a fifth to the
database as a whole.

\emph{Verbatim evidence.}  Every numeric value must come with a quote
that occurs in the paper.  The quote and the paper are put through the
same normalisation before they are compared: a compatibility fold
(which matters because the conversion renders mathematics in Unicode
italic while the model writes plain letters), ligature expansion,
case folding, the three dashes folded onto one, the rejoining of words
split across a line break, and the collapsing of punctuation and
whitespace.  A quote shorter than 24 characters is not accepted on a
substring match alone, because ``NA'' or ``Tc 200 K'' would match any
paper that happens to print that number; a second pass compares the
two with all whitespace removed, which catches the tokens the
conversion breaks across lines.  The two labels that are not numbers,
the pairing symmetry and the mechanism, are held to the same rule,
because they are claims about the compound too.  The accepted quote is
stored with the value.

\emph{Material binding.}  A value must occur in a sentence that names
the material, directly, by an alias, or through the surrounding text.
This is what keeps ``YBa$_2$Cu$_3$O$_7$ = 39~K'', which is MgB$_2$'s
number, out of the database.

\emph{How the value was obtained.}  The wording of the sentence that
carries the value is read to learn whether the number was measured or
computed.  Wording of prediction sends the value out, wording of
observation keeps it.  The reason is logged rather than the value
corrected, so that the decision can be audited against the source.
Where the wording settles nothing, the value stays and the record keeps
the extraction model's own judgement of what the number is; that
judgement, and not the wording, is what the flag in the database
records.

\emph{Physical plausibility.}  A transition temperature outside the
range its family can reach is rejected: 200~K for everything but the
hydrides, which reach further, and a tighter bound for the
single-element superconductors, whose transition temperatures are
known well enough to catch an obvious mismatch.  The check is
deliberately conservative, since rejecting a real measurement costs
one record and accepting a misattributed one corrupts the database
silently.

\emph{Cross-paper agreement.}  A compound reported by several papers
carries several values, and no check against a single paper can catch
a measurement that is the paper's own but wrong for that material.
Values within 1~K of one another are collapsed onto their cluster;
where no cluster dominates, the better-documented record is kept.
After that, rows that describe the same compound and the same
transition temperature are merged into one, taking every field that
any of them carries.  Where the merged rows disagree about how the
number was obtained, the record the sources report as an observation
survives.  A computed value for a compound the corpus measures
elsewhere is dropped: the measurement is already the database's entry
for that compound, and the computed value is a second, usually higher,
statement about it.  The two rules left 3{,}137 records where the
deduplication had produced 3{,}185, and 185 of them flagged as
computed where there had been 233 under the same rule and 261 under
the previous one.

\subsection{The audits}

The audit sample was drawn from the corpus by stratifying on the
material family and on the transition-temperature band, with the
allocation in each cell in proportion to the corpus, so that the rates
the audit gives are the rates of the database rather than of a draw
that over-weighs the rare families: iron-based materials are 23.0\% of
the sample against 23.4\% of the corpus, cuprates 21.2\% against 21.7\%,
and hydrides 10.4\% against 10.6\%.  Every cell that is not empty is in
the sample, the compositions the pipeline already knew to be difficult
are marked and reported apart, and the draw is seeded, so the same
sample can be drawn again.  All 500 of the sampled records are still
in the corpus, so the rate the judges establish is a rate of the
database itself.  Five hundred records went to each of the two judges.

The first judge was DeepSeek-V4.1-Flash, reached over its own API, a
different family from the model that did the extraction.  It was given
the composition, the stored value, and the whole of the paper the value
came from --- the conversion of the published PDF, which is what the
extraction itself read --- and it answered three questions: is the
value in the text, is the text about this material, and is the value a
measurement or a calculation.  The second judge was DeepSeek-V4-Pro,
the larger model of the same provider, given the same paper and asked
the same three questions.  The whole paper is what makes the answers
mean something: a judge that sees only the sentence that the value came
from cannot tell whether the material is named further on, and it
rejects records that are right.  What the judges see is still the
converted text and not the typeset paper, which is a limit on what a
verdict of this kind can establish, and is taken up in the discussion.

One record was removed by hand on the same principle, before the rule
above was written: its sentence said a transition temperature
\emph{could attain} 300~K, a projection rather than a value, and
storing it as one put a cuprate above every measured cuprate.

Rates are given with 95\% Wilson score intervals, which behave
sensibly at the small sample sizes involved.

\subsection{Software}

The extraction, the gates, the consolidation and the audit are Python;
the figures are drawn from the frozen corpus snapshot with
matplotlib.  Every number quoted in this paper is produced by a script
that reads the files it describes, rather than by hand.

\section{The database in use}

The database is public, and so is the machinery around it: a website
(\url{http://isspfml.cn}) that serves the materials by family and by
transition temperature, and an export that hands over the corpus
materials and the SuperCon dataset as CSV, JSON or
JSON Lines, one file per family.
The machinery is open as well: the extraction pipeline, the gates and
the audit scripts are at \url{https://github.com/isspwjlu/SCDB}.
The site is also a way in: a reader can send the PDF of a paper the
corpus does not have, and the records read out of it join the corpus
after a reviewer has looked at them.
The SuperCon companion set \cite{supercon} carries the curated
transition temperatures of the NIMS database, and its nickelate part
comes from our database.

The site therefore serves 15{,}408 records, of which 15{,}223 (98.8\%)
are measured transition temperatures and 185 (1.2\%) are values their
source reports as computed; every row says which of the two it is.  The
two proportions have to be read apart, because they come from two
different corpora.  The site's is high because more than half of what
it serves is the hand-curated NIMS catalogue, in which every row is a
measurement and there is no flag to set; the 3{,}137 records extracted
here are 94.1\% measured, and the audit puts the share of computed
values among them at 7.6\%.

The corpus is also training data.  Predicting $T_c$ from a chemical
formula is a long-standing task \cite{adatadrivens2018,machinelearn2018}; it has been done with a range of machine-learning methods, gradient boosting and random forests among them
\cite{machinelearn2023,machinelearn2024,predictingsu2024,predictingth2022,
predictivemo2020}, and the SuperconPred toolkit that this group built
on the same lines turns a formula into 45 features and predicts with
random forests, gradient boosting or Gaussian process regression
\cite{machinelearn2026}.  The point of an audited database for that use
is not only that it is larger than a hand-curated one, but that a model
trained on it can be shown where each of its training values came
from.  A model that has learned from a misattributed $T_c$ has learned
a chemical fact that is not in the paper, and one that has learned from
a computed $T_c$ can be shown to have learned a prediction rather than
an observation.

The gates are not specific to this corpus: each of the five is written
against the sentence and the material rather than against a field of a
schema, so the same set can be applied to another corpus, or to the
databases other groups have assembled.  What does not transfer is the
answer: the rates here are the rates of this corpus, and the audit that
produced them is the part that has to be repeated for another one.

\section{Discussion}

The gates remove what can be removed by looking at the text.  The
methods that mine the literature for materials have been asking for
structured output, for agreement between the model and a human, and
for a record of where each number came from
\cite{materialssyn2017,extractingac2024,fromtexttoin2025,
materialssci2024,aunifiedfram2025}; the question this paper asks is a
different one, which is how many of the numbers that survive are
right, and the answer for this corpus is that 93.4\% of the audited
records have both conditions satisfied.  The audit then measures the
other half of that answer: of the values that survive, how many are
predictions rather than observations, which in the sample is 7.6\%.  What the gates cannot do is answer the paper's own question of
how a value was obtained, because it is not a failure of reading: the
paper states the value correctly, for the right material, and says only
that it was computed.  Such a value is a correct record of what the
paper says;
what it is not is a measurement.  Those values sit mostly in the
hydrides, where the
predicted transition temperatures reach into the 200~K range and are
quoted in the same paragraphs as the measured ones.  Both kinds belong
in the corpus; what a database owes its readers is to say which is
which, and to keep the sentence that settles it.

The way forward is not to look harder at the same sentence but to change
what is required of it.  The check on how the value was obtained
currently looks for wording near the number, and it resolves a sentence
that carries both kinds of wording in favour of the observation: a
sentence that puts ``calculated $T_c$ values'' next to ``the measured
ones'' is marked as measured, and so is one that writes ``we report the
$T_c$ to be almost the same in the two phases and it reaches its peak
value of 34.6~K''; two of the fifteen missed records were stored that
way.  It also fails when the paper states the number in a table of
predicted structures, where there is no wording to read.  A check that
requires the sentence
to state how the number was obtained (measured, or predicted), and that
gives the wording of prediction precedence over the wording of
observation, would mark the SnH$_{12}$ row and the YH$_6$ range as what
they are, and
would also have caught the range that was stored as a point.

Three limits of the audit itself should be stated.  Both judges are
models, not readers: they see the converted text of the paper rather
than the paper.  They are given the whole paper, and they still
disagree with each other on twenty records out of 500 --- so the
agreement between them is a fact about the two models and not a
guarantee about the records.  And the sample is a sample: it is drawn
in proportion to the corpus, so that its rates are the corpus's rates
up to the sampling error, and every rate quoted here carries the
interval of that error.

The two readings of the same records agree on the level, which is the
point of drawing the sample in proportion to the corpus: the flag
marks 5.9\% of the corpus as not experimental, and the audit, reading
the same sentences afresh, finds 7.6\%.  The two do not agree
exactly, and the difference is the price of a flag that is set once,
from one reading of how the source paper frames the value: it catches
23 of the 38 computed values the judges found and misses the other 15,
none of which it mislabels in the other direction.  So the corpus's
5.9\% is a lower bound and the audit's 7.6\% the better estimate, and
the two leave the true share between them.  What they do agree on
exactly is which records these are: the 23 the flag marks as computed
are the 23 the judges mark as computed, so a reader who wants the
computed values can ask for them by that mark and get the right ones
out.  Both readings also place them in the high-$T_c$ tail and among
the hydrides, which is where the distinction between the two kinds
matters most.

A second limit is the corpus itself.  The papers are the ones
that could be fetched, and the databases of the field are the ones
that were not: the NIMS SuperCon set \cite{supercon}, the curated
superconductor corpus of Yamaguchi and co-workers
\cite{superconduct2022}, the automatic extraction work of Foppiano and
co-workers \cite{automaticext2023}, and the recently assembled
dataset of superconductors with their crystal structures
\cite{3dscadataset2023}.  The corpus of this work is a complement to
those rather than a replacement: it is smaller, and every record in it
can be checked against a sentence.

Set against the previous work, the difference is one of kind.  The
databases assembled with LLMs are checked, where they are checked at
all, by asking another model to read the same sentences
\cite{thenortheast2025}, and the extraction of superconductors from
the literature has been addressed as a corpus-building problem
\cite{automaticext2023,superconduct2022} rather than as a question of
what survives the extraction.  Storing the sentence beside the value
turns that question into an answerable one, and the answer here is
that a database assembled this way is usable, and that it is usable
because the records that carry a computed value are marked as such.
Marking them is cheap: it is a field of the record, and the sentence
the value came from is there to check it against.

The question of how a value was obtained is not peculiar to
superconductors, and the ways of handling it are not either: the
hallucinations of language models are now catalogued
\cite{largelanguag2026}, the agents that read materials literature
are being given the same kind of task \cite{modularlarge2026,
fromdatatoth2026}, and the tools that put a person in the loop for
the extraction are being built \cite{scidasynthin2025}.  What the
superconductivity case adds is a quantity that is easy to check and
easy to get wrong at the same time.  The hydrides are the sharp
case.  Their predicted transition temperatures reach into the 200~K
range \cite{conventional2015,superconduct2019}, they are quoted in the
same paragraphs as the measured ones, and they are what a search for
a high-$T_c$ superconductor would return first.

Two things make this a question for a physics journal rather than one
about data handling.  The first is what the answer is about: a
predicted and a measured transition temperature answer different
questions, and a corpus that does not say which is which cannot answer
either; the hydrides are where the two kinds sit closest together.
The second is that the answer is a property of the literature rather
than of one corpus: the gates and the audit are written against the
sentences, and can be pointed at any set of superconducting papers
without changing what they measure.

\section*{Data availability}

The material records, with the sentence each value was taken from, the
source DOIs and the derived fields, are available from the project
website (\url{http://isspfml.cn}) together with the export described
above, in CSV, JSON and
JSON Lines.  The corpus snapshot used in this paper (26{,}210 papers,
3{,}137 material records over 1{,}371 compositions) is the 21 November
2026 release.

\section*{Code availability}

The extraction pipeline, the gates, the consolidation pass and the
scripts that read the audit verdicts are released at
\url{https://github.com/isspwjlu/SCDB}.  The datasets are not in
that repository: the SuperCon catalogue is distributed by NIMS for
research use, and the material records are the database the website
serves.  The audit sample and the verdicts of the two judges are
therefore not there either, though the
scripts that consume them are, so that the numbers quoted here can be
recomputed from the same files once the data is in hand.

\section*{Author contributions}

W.J.L. designed the extraction and the verification protocol, wrote
the code, produced the statistics and the figures, and wrote this
paper.

\section*{AI use statement}

The corpus was extracted with MiniMax-M3.1-Flash, and the independent
audit of the extracted records was performed with DeepSeek-V4.1-Flash
and DeepSeek-V4-Pro; these models were used for the purposes described
in the text and for nothing else.  Claude Code was used for writing the
extraction, verification and figure code, for producing the figures in
this paper, and for drafting and revising the manuscript; the numbers
reported here were computed by the released scripts, and the
interpretation of them is the author's.

\section*{Acknowledgments}

The author thanks Claude Code for the programming, the figures, the
writing of this manuscript and the website, and the developers of
docling \cite{doclingconve2026} for the document conversion the pipeline
depends on.

\section*{Competing interests}

The author declares no competing interests.

\appendix
\section{The database as a website}

The database is not only a table; it is a site that lets a reader use
it without writing code, and it is the form in which most readers meet
the material described in this paper.  It lives at
\url{http://isspfml.cn}.  Figure~\ref{fig:site} shows the two pages
that carry it.

The \emph{materials} page (Fig.~\ref{fig:site}b) is the entry point.
The records are listed with their transition temperature, their family
and their mechanism, and they can be narrowed by a search on the
composition, by the family, by a range of transition temperatures and
by a sort order.  The families in the menu are the ones the extraction
actually produced, with their counts, so a reader who wants the
nickelates sees how many of them there are before asking for them.
Clicking a record opens it: the page carries the composition and the
transition temperature at the top, the family and the DOI of the source
paper, the origin of the row, and then the properties in the groups the
extraction filled --- the base superconducting quantities, the pressure
dependence, the isotope effect, the specific heat, the Fermi
parameters, the crystal structure and the preparation conditions.  A
property the paper did not state is left empty rather than filled with
a default, which is the rule the gates apply to the values themselves.

The \emph{home} page (Fig.~\ref{fig:site}a) gives the size of the
corpus, its highest record and the families it holds, with the five
highest transition temperatures below them.

The \emph{charts} page gives the distribution of transition
temperatures and the composition of the families, with the four
relations of Fig.~\ref{fig:relations} beside them.  The figures of this
paper are these charts, redrawn.

The \emph{download} page is the way out.  The corpus materials, the
SuperCon companion set are offered as CSV, JSON or
JSON Lines, one file per family or one archive for several, and every
one of them carries the sentence each value was taken from, so that a
reader who wants to check a value against its source can do so without
writing anything.  The pages are public; the export asks for a
registered account, so that the number of downloads can be counted.

The \emph{upload} page is the way in.  A registered reader can send
the PDF of a paper, and the file is read on the machine that holds the
corpus, by the same pipeline that read the corpus and behind the same
gates: the value has to be in the text, it has to belong to the
material, and how it was obtained has to be recorded.  What
the extraction found is shown to a reviewer --- the compositions, the
transition temperatures, and the sentence each value was taken from ---
and what the reviewer approves joins the corpus, with the paper it came
from and its origin marked as a contribution.  A record that arrives
this way is held to the same standard as one that came out of the
corpus sweep, and the counts of this path are counted apart from the
counts of that one, because they are a different population.  The corpus
this paper describes is the release of 21 November 2026 and predates
this path, so nothing has been admitted through it yet.

The site is live and the extraction continues, so the counts it shows
grow; Figure~\ref{fig:site} was made on the release this paper
describes.

\bibliographystyle{apsrev4-2}
\bibliography{scdb_main}

\end{document}